\documentclass[12pt]{iopart}
\usepackage{graphicx}  
\begin{document}

\title{Using Cold Atoms to Measure Neutrino Mass}

\author{M. Jerkins}
\address{Center for Nonlinear Dynamics, University of Texas, Austin, TX 78712}
\author{J. R. Klein}                                                           
\address{Department of Physics, University of Pennsylvania, Philadelphia, PA, 19104}                                                                          
\author{J. H. Majors}                                                          
\address{Center for Nonlinear Dynamics, University of Texas, Austin, TX 78712}
\author{F. Robicheaux}                                                         
\address{Department of Physics, Auburn University, Auburn, AL 36849}           
\author{M. G. Raizen}                                                          
\address{Center for Nonlinear Dynamics, University of Texas, Austin, TX 78712}

\begin{abstract}
We propose a $\beta$-decay experiment based on a sample of ultracold atomic tritium.  These initial conditions enable detection of the helium ion in coincidence with the $\beta$.  We construct a two-dimensional fit incorporating both the shape of the $\beta$-spectrum and the direct reconstruction of the neutrino mass peak.  We present simulation results of the feasible limits on the neutrino mass achievable in this new type of tritium $\beta$-decay experiment.

\end{abstract}

\maketitle

\section{Introduction \label{sec:intro}}
	The past decade has transformed our understanding of the
neutrino; nevertheless, the absolute scale of the neutrino mass
remains unknown.  The best neutrino mass limits from direct
measurements come from the tritium endpoint experiments Mainz and
Troitsk~\cite{mainz,troitsk}, both of which place $m_{\nu} < 2.2$~eV.
Measurements of the cosmic microwave background, coupled with
cosmological models, have led to somewhat better (but model-dependent)
constraints of $m_{\nu} < 1$~eV~\cite{wmap}.

        The next generation of tritium endpoint measurement is now
        being pursued by the KATRIN experiment~\cite{KATRIN}.  They
        expect to push the limit on the neutrino mass as low as
        $m_{\nu} < 0.2$~eV.  An independent avenue of research is
        neutrinoless double $\beta$-decay, which could test the
        Majorana nature of the neutrino and possibly determine its
        mass~\cite{neutrinoless}.

     We propose here a new approach, fundamentally different
     from both KATRIN and neutrinoless double $\beta$-decay.  Our work
     is motivated by the recent development of general methods for
     trapping and cooling of atoms, which enable the creation of a
     sample of ultracold atomic tritium.  We first describe the atomic
     trapping and cooling methods and then outline a prototype of a
     neutrino mass experiment.  We present detailed simulation results
     and the detector requirements necessary to reach sub-eV
     sensitivity on the neutrino mass.

\section{Slowing and Cooling Methods}

     Cooling of atomic translational motion has been the topic of
     intense research for the past thirty years.  The standard
     approach to date is laser cooling~\cite{metcalf}, which has been
     applied to cooling and trapping of radioactive alkali atoms in
     order to probe the weak interaction~\cite{freedman, behr, vetter, fang}.
     Despite the enormous success of this method, it has been limited
     to a small set of atoms due to the requirement of a cycling
     transition that is accessible with lasers.  In particular,
     hydrogenic atoms have not been amenable to laser cooling.
     Trapping and cooling of hydrogen atoms was accomplished in a
     dilution refrigerator, followed by evaporative cooling, but these
     methods have not been extended to other isotopes of
     hydrogen~\cite{hess}.

     Over the past few years a more general method has been demonstrated
     in a series of
     experiments.  The starting point is the supersonic molecular
     nozzle which creates a very monochromatic but fast
     beam~\cite{Scoles}.  Paramagnetic atoms or molecules are seeded
     into the beam in a region of high density and decouple from the
     carrier gas downstream.  These atoms are stopped with a series of
     pulsed electromagnetic coils, an ``atomic coilgun.''  Such a
     device has been used to stop a beam of metastable neon, molecular
     oxygen, and atomic
     hydrogen~\cite{Narevicius07,Narevicius08,Narevicius08_2,merkt07,merkt07_2,merkt08}.
     Once the atoms are magnetically trapped, they can be further
     cooled using a method of single-photon cooling, which is based on
     a one-way barrier~\cite{Price08,bannerman}.  Together the atomic
     coilgun and single-photon cooling provide a general two-step
     solution to the trapping and cooling of paramagnetic atoms or
     molecules.  In particular, these methods will work well on atomic
     tritium, which has a suitable 12.3 year half-life.

\section{Prototype Experiment}

     We consider an experiment to observe the $\beta$ decay of
     ultracold atomic tritium.  The decay produces an outgoing
     $^3$He$^+$ ion and a $\beta$, both of which can be detected.  We
     need a spectrometer to measure the energy of the $\beta$, along
     with a non-invasive technique for measuring two components of its
     momentum.  By utilizing the coincidence between the $\beta$ and
     the $^3$He$^+$ ion, we can determine the ion's three momentum
     components from its time-of-flight.  Measurement of the
     four-momenta of the ion ($\tilde{p}_{\rm He}$) and the $\beta$
     ($\tilde{p}_{\beta}$) yields the neutrino mass squared:
\begin{equation}
m^2_{\nu} = \tilde{p}_{\nu} \cdot \tilde{p}_{\nu} =
(\tilde{p}_{^3H}-\tilde{p}_{He^3}-\tilde{p}_{\beta}) \cdot
(\tilde{p}_{^3H}-\tilde{p}_{He^3}-\tilde{p}_{\beta})
\end{equation}

     The advantages of this approach include: an extremely thin source that results in low scattering; an atomic tritium source with simple final state effects; a coincidence measurement with the $\beta$ to reduce background; a direct
     neutrino mass peak reconstruction; and the utilization of at
     least 500~eV of the $\beta$ energy spectrum.  Nevertheless,
     this approach faces several experimental challenges,
     particularly regarding the measurement of the $\beta$ momentum
     to sufficient precision, and trapping enough tritium atoms to
     obtain sufficient statistics.
      
     We address these challenges with a proposed experimental setup
     that would consist of three detectors shown in Figure 1: a
     microchannel plate (MCP) to detect the helium ion, a
     spectrometer to measure the $\beta$'s energy, and an optical
     lattice of rubidium Rydberg atoms capable of measuring two of
     the $\beta$'s three momentum components.

\begin{figure}[h]
\begin{center}
{{\resizebox{3.5in}{!}{\includegraphics{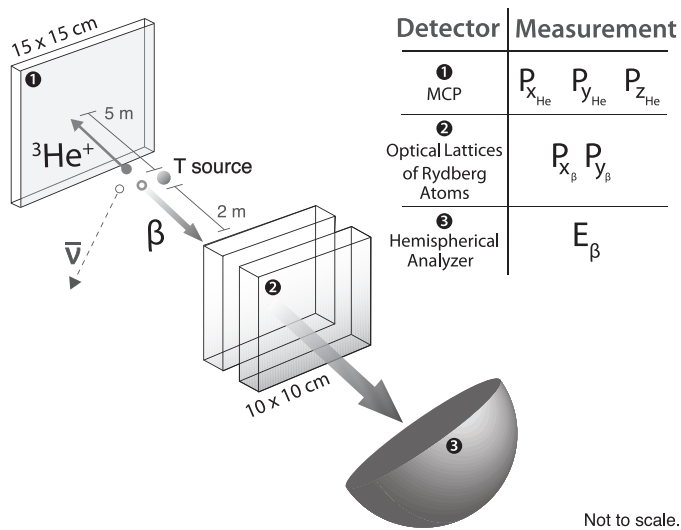}}}}
\caption{Experimental setup of the three detectors proposed for
kinematic reconstruction of the neutrino mass: a microchannel plate
(MCP), optical lattices of rubidium Rydberg atoms, and a
spectrometer.}
\end{center}
\end{figure}

	We can place the $\beta$-spectrometer close to the source,
	with the MCP for the $^3$He$^+$ ion detection several meters
	away from the source.  Using the $\beta$ event detected by the
	spectrometer as the initial time, we can determine the
	time-of-flight of the ion to the MCP.  Combining the
	time-of-flight with the MCP hit position yields the three
	momentum components of the helium ion.  For example:
\begin{equation}
p_{x} = {\gamma}mvsin{\theta}cos{\phi}
\end{equation}
where $v = z/(TOF)cos{\theta}$ and $\theta$ and $\phi$ are
reconstructed from the MCP hit position assuming the tritium decay
came from the center of the source. Here $z$ is the distance from the MCP
to the source.

      The background event rate from the MCP would be
      $<1$~event/cm$^2$/s~\cite{siegmund}, where cosmic ray
      events are eliminated either by deploying the detector in
      an underground laboratory or by implementing an efficient veto.
      Although the coincidence in the $\beta$-spectrometer would be
      helpful, for any given $\beta$ event of the correct energy there
      will be a $7\%$ chance of seeing a background MCP hit, given
      that the coincidence time between the $\beta$ and the ion will
      be on the order of 0.3~ms.  In order to evaluate our ability to
      discriminate true events from backgrounds, we simulated data in
      which the MCP hit position was randomized, and we studied how
      our reconstruction algorithm evaluated the neutrino mass squared for
      such random events.  Such events typically reconstruct to be
      more negative than $-10^{6}$~eV$^{2}$ and would be clearly
      separated from true helium ion hits.  Our simulations indicate
      it is possible to reduce backgrounds to
      $1.0$x$10^{-5}$, not including the rejection due to the
      coincidence requirement, simply by cutting any events that
      reconstruct the neutrino mass squared to be more negative than
      $-5000$~eV$^{2}$.  This cut introduces negligible bias into the neutrino mass squared peak.

        In order to measure the momentum of the $\beta$ without
        significantly altering its energy, we propose exploiting the
        effect of a passing electron on Rydberg atoms~\cite{kleppner,gallagher}.
        In the $\beta$'s flight path before it reaches the
        spectrometer, we create an optical lattice filled with
        rubidium atoms in the ground state~\cite{gould_2006,
        raithel_2000}.  Using laser excitation, we can excite the
        atoms to a high Rydberg state~\cite{saffman_2008,
        raithel_2005}, such as 53s.  When the $\beta$ passes one of these atoms, it
        can excite the atom from a 53s state to a 53p state,
        and the atom will remain trapped in its optical
        lattice position.  We propose slowing the electrons with a
        controlled voltage soon after they leave the source so that by
        the time they reach the optical lattice, they have a maximum
        energy of 900~eV, which increases their cross section for
        exciting a Rydberg atom to 0.36 x 10$^{-9}$ cm$^2$\footnote{We calculated this transition cross section using the first order Born approximation, which is applicable because the electron energy is more than 10$^7$ times larger than the transition energy, and the transition is dipole allowed.  We numerically computed the radial part of the transition matrix element by using a Numerov algorithm to compute the radial orbitals on a square root mesh in $r$.  We numerically integrated the radial orbitals times the Bessel function, $j_{1}(qr)$, for the transition operator using 4th order integration.  To obtain the total cross section, we numerically integrated over the momentum transfer $q$ from $q_{min} = k - \sqrt{k^2 - 2{\Delta}E}$ to a $q_{max} = 0.25/n$ using equally spaced points in $q$ with a ${\Delta}q$ = 0.01/$n^2$.}. When a
        $\beta$ signal is detected downstream in the spectrometer, the
        53s atoms are optically de-excited using
        STIRAP (stimulated Raman adiabatic passage)~\cite{raithel_2005}, and an electric field of 100 V/cm
        is ramped within $\sim$130~ns to ionize any Rydberg atoms in a
        53p state. Once the atoms are ionized, they will be detected
        by a multi-hit position-sensitive MCP.  Based on realistic density limits, the $\beta$
        will excite several Rydberg atoms as it passes through the
        optical lattice, so we will be able to obtain the projection
        of a track from the passing $\beta$.

      In order to obtain the two $\beta$ momentum components necessary
      for reconstruction, we need to have a second optical lattice to
      project the momentum component in a direction orthogonal to the
      first.  By combining the track projections from these two MCPs
      with the energy measurement from the spectrometer, we can
      reconstruct the momentum of the $\beta$ that traversed the
      optical lattices using equation (2) and the reconstructed
      velocity:
\begin{equation}
v = c(1 - 1/(T/m + 1)^{2})^{1/2}
\end{equation}
where $T$ is the kinetic energy of the $\beta$ as measured in the
spectrometer and $\theta$ and $\phi$ are obtained from the $\beta$
tracks in the optical lattices.  Using Rydberg atoms with a principal
quantum number n=53 would result in a negligible change in the
$\beta$'s four-momentum as it passes.  We estimate that we can obtain
a density of 10$^{11}$ atoms/cm$^{3}$ in the optical
lattice~\cite{metcalf}, and we expect the passing $\beta$ to excite an
atom within 5 $\mu$m, leading to a high spatial resolution.

       The two major sources of backgrounds that must be eliminated
       for this Rydberg technique are collisions and black body
       excitations.  Holding the Rydberg atoms in an optical lattice
       eliminates collisions that could cause spurious transitions to
       the 53p state~\cite{gould_2006}.  By surrounding the optical
       lattice with a wire mesh, we can eliminate most of the black
       body radiation that could excite atoms from the 53s to the 53p
       state.  The spacing of the mesh would be small compared to the
       microwave wavelength, suppressing blackbody emission of the
       mesh itself.  Additionally, the rubidium atoms can be
       periodically cycled back to the ground state and then up to the
       Rydberg 53s state~\cite{saffman_2008, raithel_2005}, which will
       prevent background 53p events from accumulating, while still
       allowing the atoms to spend most of their time in the 53s
       state.  This non-invasive method may find other applications in
       the detection of low-energy electrons.

\section{Simulation Results}

        Our current experimental simulation makes several assumptions
        about detector precision in order to determine the required
        equipment.  We assume an MCP of 15~cm x 15~cm with a timing
        resolution of 20~ps and a high spatial resolution of
        2~$\mu$m~\cite{mcp2009,mcp2008,mcp}. It is placed 5~m from the
        tritium source and has a 44$\%$ efficiency for detecting an
        ion when it is hit.  The tritium source is modeled as a
        100~$\mu$m sphere at a temperature of 1~$\mu$K.  Given that
        the density of the source cannot exceed 10$^{15}$
        atoms/cm$^{3}$ and that the radius of the source is 50~$\mu$m,
        the column density of the source is less than 10$^{13}$
        atoms/cm$^2$.  We therefore estimate multiple scattering
        within the source to be small and do not include it in the
        simulation.  The $\beta$-spectrometer is a hemispherical
        analyzer with an energy resolution of 5~meV, which is
        reasonable given current devices~\cite{spectrometer}.
        Simulations indicate that the Rydberg atom method of measuring
        the $\beta$ momentum results in a resolution that varies from
        40~meV/c to 2.8~eV/c depending on the $\beta$'s
        four-momentum. We assume a large Rydberg atom optical lattice
        with dimensions 10 cm x 10 cm x 1 cm placed 2 m from the
        source, which optimizes the detector's resolution and
        solid-angle acceptance.

     Our simulated $\beta$ spectrum includes first-order final state
     corrections.  In tritium $\beta$-decay, the helium ion is formed
     in the ground state in 70$\%$ of the decays, and our simulation
     simplifies the true spectrum of final states by assuming that the
     helium ion goes into the first excited state for the remaining
     30$\%$ of the decays.  For more than 99.9$\%$ of the events, the
     magnitude of the reconstructed neutrino mass is larger when the
     wrong state is assumed for the helium ion, which provides us with
     a simple method of determining the true state of the helium ion.
     This method does not bias the neutrino mass fit in any significant way.

       Both the neutrino's reconstructed mass peak and the shape of
       its $\beta$-spectrum contain information about its mass.  In
       order to utilize all of this information, we perform a
       maximum-likelihood fit using two-dimensional probability
       density function (pdf).  We create a series of 2D pdfs using
       much higher statistics than we use for our simulated data.
       Each of the six pdfs we create has a different assumed neutrino
       mass, and the assumed mass values are $4.0$~eV apart.  Figure 2
       shows the 2D pdf for the case of zero neutrino mass.  By
       interpolating between the pdfs, we find the most likely value
       for the neutrino mass for a particular data set.

\begin{figure}[h]
\begin{center}
{\rotatebox{0}{\resizebox{3.7in}{!}{\includegraphics{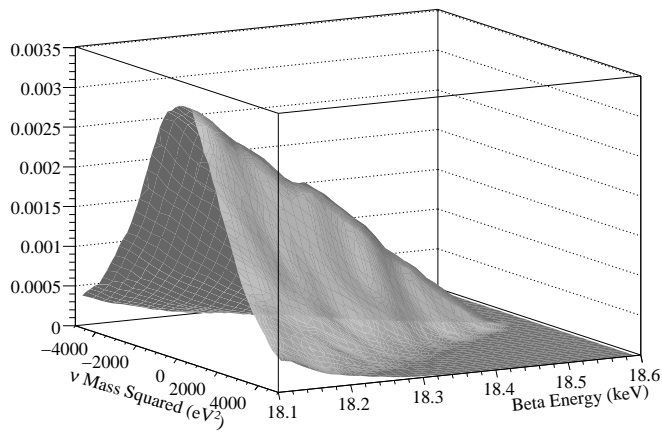}}}}
\caption{One of the six 2D probability distribution functions used in
the fitting process.  This sheet corresponds to a neutrino mass of
0.0~eV, and the data set was fit by interpolating between pdfs of
different assumed neutrino mass.}
\end{center}
\end{figure}

\begin{figure}[h]
\begin{center}
{\resizebox{3.5in}{!}{\includegraphics{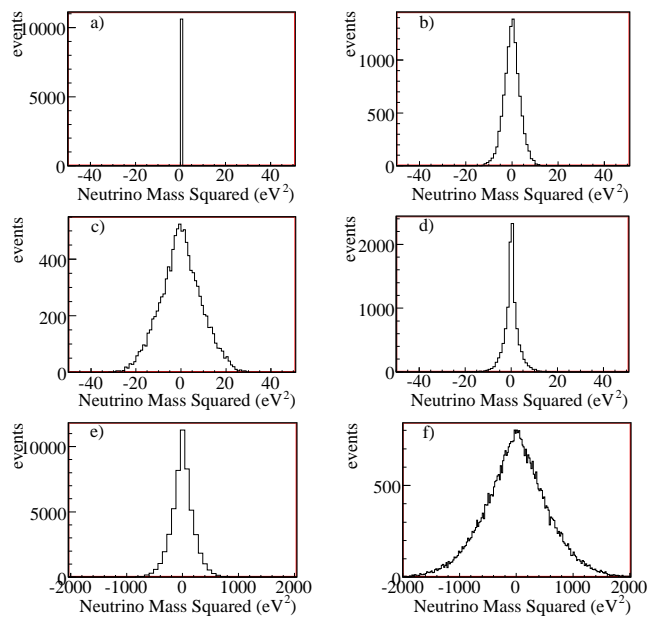}}}
\caption{Reconstructed neutrino mass squared peak broadenings caused
by various uncertainties and detector resolutions.  a) All smearings
turned off.  b) $\beta$ energy resolution.  c) $^{3}$He ion's MCP
binning resolution.  d) $^{3}$He ion's MCP timing resolution.  e)
$\beta$ momentum resolution. f) $^{3}$H 1$\mu$K initial temperature.}
\end{center}
\end{figure}

       Unlike previous tritium $\beta$-decay experiments that utilize
       information only a few eV away from the endpoint, our fit
       extends back to 18.1~keV, a full 500~eV from the endpoint.  The
       statistics gained by moving away from the endpoint
       substantially improve the precision on the neutrino mass even
       as the spread in reconstructed mass gets broader.  Figure 3
       shows how individual detector and reconstruction uncertainties
       contribute to broadening the reconstructed neutrino mass
       squared peak, especially the $\beta$ momentum measurement and
       the initial $^3$H temperature.  These smearings create large
       uncertainties for each reconstructed event, but the uncertainty
       in the mean of the peak decreases with added statistics.
       Combining this neutrino mass peak information with the
       information from the beta spectrum fit allows for a sub-eV
       determination of the neutrino mass.  Clearly, systematic shifts
       in the mean of the reconstructed mass spectrum would have to be
       controlled at a very high level, but calibrations of the
       spectrometer using the conversion electron from $^{83m}$Kr as
       well as information from the energy spectrum itself should
       allow us to mitigate these effects.

\begin{figure}[h]
\begin{center}
{\resizebox{3.5in}{!}{\includegraphics{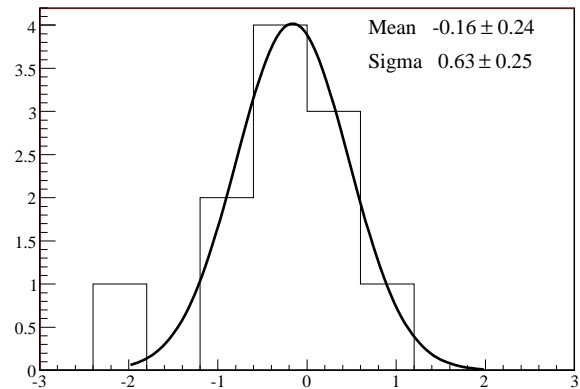}}}
\caption{Pull distribution comparing the fit results shown in Table 1
to the neutrino mass that was assumed in the various simulation
trials.}
\end{center}
\end{figure}

\begin{figure}[h]
\begin{center}
{\resizebox{3.5in}{!}{\includegraphics{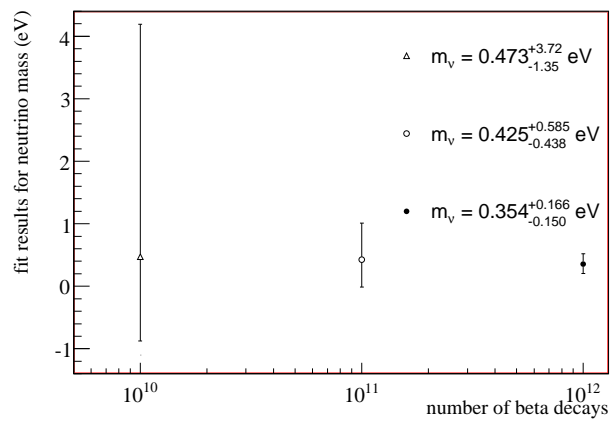}}}
\caption{MINUIT fit results and MINOS errors from simulated data runs
in which the neutrino mass was 0.4~eV.}
\end{center}
\end{figure}

\begin{table}[h]
\centering
\begin{tabular}{| c | c | c | c |}
\hline Assumed $m_{\nu}$ & Fit $m_{\nu}$ & (+)error & (-)error \\
  \hline 0.2 & 0.239 & 0.174 & 0.153 \\ \hline 0.4 & 0.354 & 0.166 &
  0.150 \\ \hline 0.6 & 0.690 & 0.270 & 0.203 \\ \hline 0.8 & 0.794 &
  0.247 & 0.215 \\ \hline 1.0 & 0.813 & 0.246 & 0.207 \\ \hline 5.0 &
  5.188 & 0.402 & 0.378 \\ \hline
\end{tabular}
\caption{MINUIT fit results and MINOS errors for simulated data runs
which had different assumed neutrino masses.}
\end{table}

       In order to reach an $m_{\nu}$ limit comparable to KATRIN's, on
       the order of $10^{12}$ tritium decays would have to occur,
       which corresponds to trapping ~$2$x$10^{13}$ tritium atoms as a
       source if the experimental live runtime is $75\%$ of one year.
       That many atoms cannot be contained in a single $100~\mu$m
       diameter trap, which cannot have a density exceeding $10^{15}$
       atoms/cm$^3$ without contributing significant scattering in the
       source.  Any feasible experiment, therefore, will require an
       array of tritium traps spaced far enough apart to allow the fit
       reconstruction to accurately determine the decay origin.  A
       third optical lattice filled with Rydberg atoms could also be
       used to detect a track from the beta as it leaves the source,
       aiding in the reconstruction of where the decay occured in the
       extended source.  Tritium sources can be stacked by repeated launching and trapping.  The primary limitation to the number that can be stacked is the trap lifetime.  This trap lifetime can be on the order of 5-10 minutes using appropriate cryogenic cold fingers and careful bake-out of the chamber.  We estimate that the necessary 10$^{13}$ tritium atoms can be accumulated in this fashion.

      Table 1 shows the results of the fit assuming $10^{12}$ tritium
      decays for six different assumed neutrino masses.  Figure 4
      shows the pull distribution of the fit results shown in Table 1,
      and its shape is consistent with a normal Gaussian. Figure 5
      indicates how the size of the fit uncertainties increases as the
      number of tritium decays decreases.

\section{Conclusions}
     Our method of investigating tritium $\beta$-decay has the
     potential to establish an interesting limit on the neutrino mass.
     Although several engineering challenges remain, such an
     experiment provides an independent and complementary method of
     measuring the neutrino mass.

We acknowledge support from the Alfred P. Sloan Foundation (JRK) and
the Department of Energy (FR, JRK).  We also acknowledge support from the
Sid W. Richardson Foundation, the State of Texas Advanced Research
Program, and the National Science Foundation (MGR).  We thank
Huaizhang Deng for helpful discussions.

\end{document}